\documentclass[10pt,letterpaper,twocolumn]{article}
\usepackage{ol3}
\usepackage[draft]{hyperref}
\usepackage{amsmath}
\begin{document}
\twocolumn[
\title{Force-insensitive optical cavity}
\author{Stephen Webster$^{*}$ and Patrick Gill}
\address{National Physical Laboratory, \\ Hampton Road, Teddington, Middlesex, TW11 0LW, UK \\ $^*$Corresponding author: stephen.webster@npl.co.uk}
\begin{abstract}
We describe a rigidly-mounted optical cavity which is insensitive to inertial forces acting in any direction and to the compressive force used to constrain it. The design is based on a cubic geometry with four supports placed symmetrically about the optical axis in a tetrahedral configuration. To measure the inertial force sensitivity, a laser is locked to the cavity while it is inverted about three orthogonal axes. The maximum acceleration sensitivity is $2.5\times 10^{-11}/\mathit{g}$ (where $\mathit{g}=9.81\,\mathrm{m}\,\mathrm{s}^{-2}$), the lowest passive sensitivity to be reported for an optical cavity.
\end{abstract}
]

\noindent Optical cavities find wide application in atomic, optical and laser physics~\cite{Brooker}, telecommunications, astronomy, gravitational wave detection~\cite{Willke} and in tests of fundamental physics~\cite{Eisele} where they are used as laser resonators, for power build-up, filtering, spectroscopy and laser stabilisation. In metrology, they are essential to the operation of optical atomic clocks, both enabling high-resolution spectroscopy and providing stability in the short-term during interrogation of the atomic transition~\cite{Gill}. In this case, a laser is stabilised to a mode of a passive optical cavity consisting of two highly reflective mirrors contacted to a glass spacer. The frequency of the laser is defined by the optical length of the cavity, thus, for spectral purity, perturbations to this length must be minimized. To this end, the cavity is both highly isolated from its surroundings~\cite{Webster2004} and designed to be insensitive to environmental fluctuations~\cite{Webster2008}. Having suppressed the effects of technical noise a fundamental limit is reached, that due to thermal noise~\cite{Numata}, and the lowest fractional frequency instabilities to have been reported are in the $10^{-16}$ region on a timescale of 0.1-100\,s~\cite{Young,Jiang}.

Several applications of optical clocks, including geodesy~\cite{Chou}, tests of fundamental physics in space~\cite{STEQUEST} and generation of ultra-stable microwaves for radar~\cite{Fortier}, will demand that the supreme performance of an ultra-stable laser be available for use in a non-laboratory environment with stabilities targeted at the $10^{-15}$ level. However, for this to be possible, a radical re-design of the cavity is required. The cavity is typically housed in a laboratory having low ambient vibration levels and is designed to be insensitive to vibrations~\cite{Millo}. It rests under its own weight and must remain in a fixed orientation with respect to gravity. For operation in micro-gravity, all degrees of freedom must be constrained and the effect of the additional forces this introduces must be taken into account. Also, outside the laboratory, it is likely that the cavity will experience larger inertial forces. Thus, the challenge is to make a cavity which is both mounted rigidly and insensitive to the forces acting upon it.

Leibrandt et al. present a design that is insensitive to both vibrations and orientation~\cite{Leibrandt1}. A spherical cavity is held rigidly at two points on its diameter and a configuration is found where its length is insensitive to the forces used to support it. The acceleration sensitivities along three orthogonal axes are $4.0/16/31\times 10^{-11}/\mathit{g}$ and, with active feed-forward correction for acceleration, they are $1.1/0.6/0.4\times 10^{-11}/\mathit{g}$~\cite{Leibrandt2}. Here, we present an alternative design based on a cubic geometry with a four-point tetrahedral support. The cavity is also insensitive to the forces used to support it and has a maximum passive acceleration sensitivity of $2.5\times 10^{-11}/\mathit{g}$.
\begin{figure}[htb]
\centerline{\includegraphics[width=8cm]{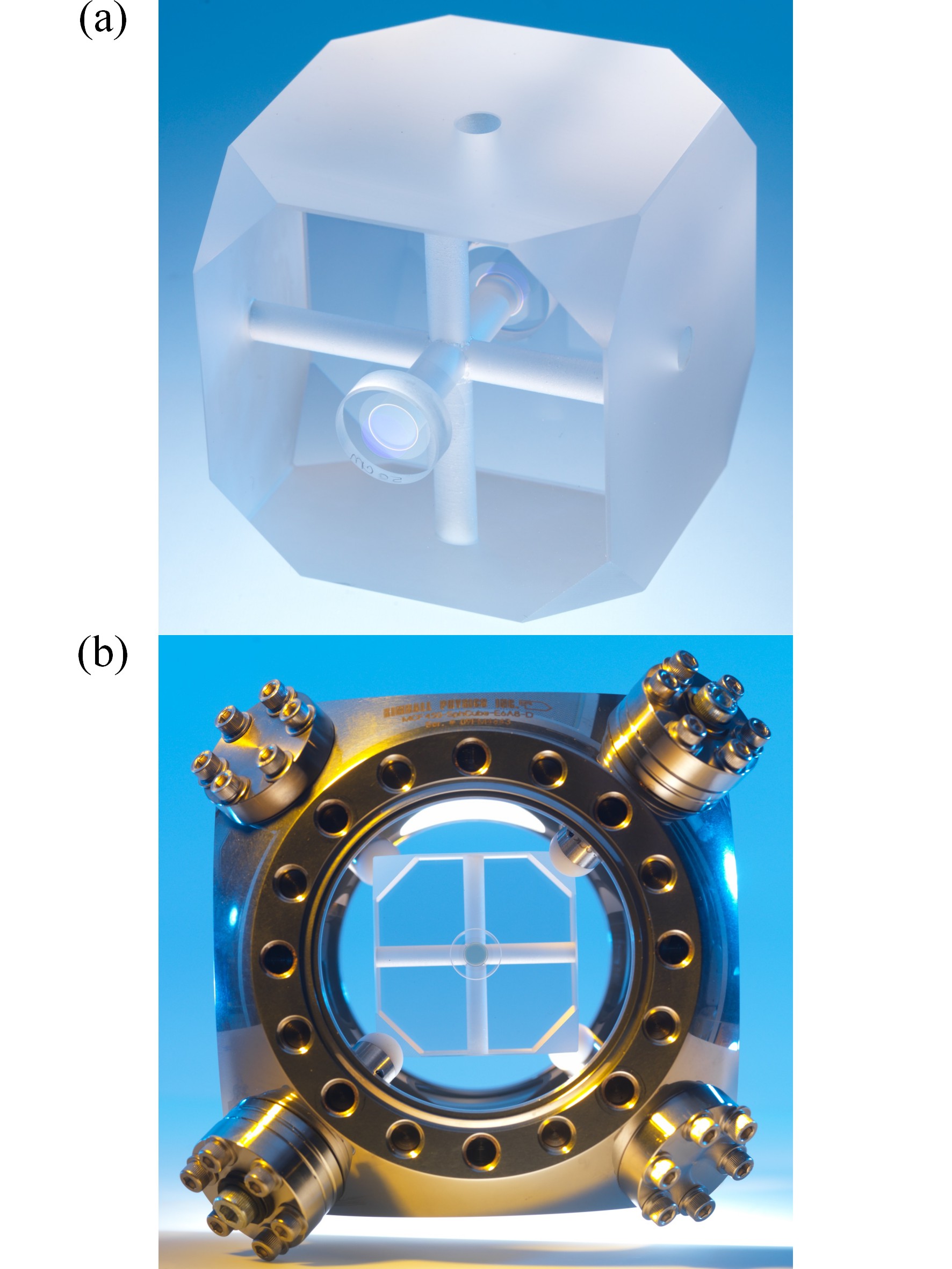}}
\caption{Cubic cavity (a) in isolation and (b) mounted within vacuum chamber. \label{fig1}}
\end{figure}

For the cavity to be insensitive to inertial forces, there must be symmetry both in the geometry of the cavity and its mount and in the forces acting at the supports. This ideal must be combined with the requirement for the cavity to be constrained in all degrees of freedom. Four identical supports in a tetrahedral arrangement constitute a sufficient and symmetric constraint for a rigid body in three dimensions and their symmetrical arrangement with respect to an axis results in a cubic geometry. The geometry is shown in Figure~\ref{fig1}. The optical axis passes through two opposing face centres of the cube and the supports are at four of the vertices of the cube which also lie at the vertices of a tetrahedron. The vertices are truncated and the support surface is spherical. This aspect of the geometry ensures that the cavity, in addition to being constrained against displacement, is also constrained against rotation.

By symmetry, the cavity is insensitive to inertial forces due to acceleration in all degrees of freedom (linear and rotational): axial displacements at the mirror centres are either zero or cancel out so that there is no net change of length on axis. A second-order sensitivity remains: that due to the inertial force arising from uniform rotation (centrifugal force). Using finite-element analysis, this is calculated to be $-5.3\,(+7.6)\times 10^{-12}\,\mathrm{s}^2$ for rotation about the optical axis (about axes perpendicular to the optical axis).

\begin{figure}[htb]
\centerline{\includegraphics[width=7.5cm]{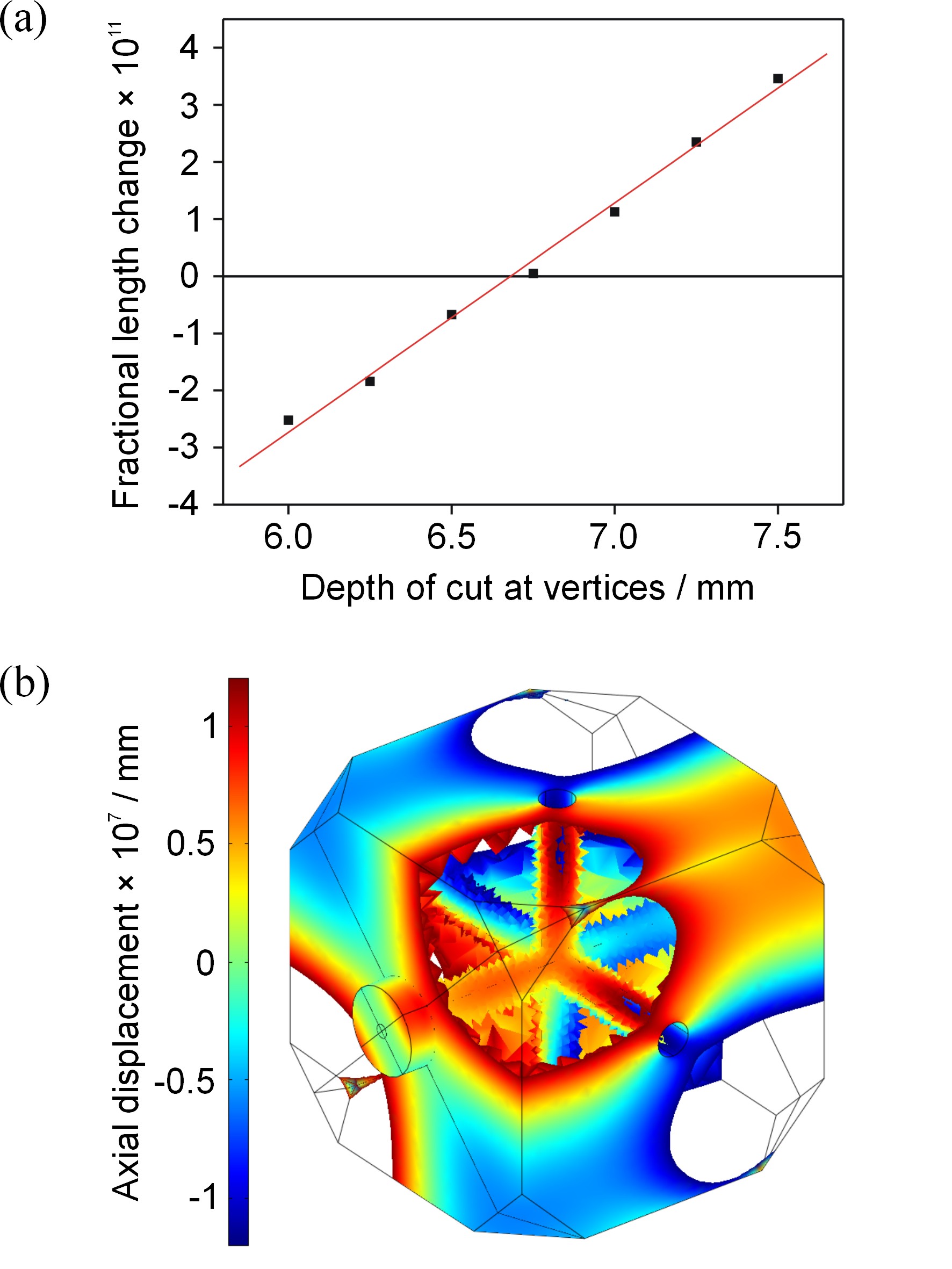}}
\caption{Results of finite-element analysis in which a compressive force of 1\,N is applied at each of the supports. (a) Fractional length change as a function of the depth of cut at the vertices. (b) Axial displacement for a cut depth of 6.7\,mm. \label{fig2}}
\end{figure}

A compressive force, directed towards the centre of the cavity, is applied at each of the supports and, in general, this causes equal and opposite axial displacements at the mirror centres and a corresponding change in length. This is undesirable in that the length is then susceptible to changes in the applied force due to, say, temperature fluctuations or mechanical relaxation. However, the truncation of the cube vertices, presents a method by which this effect can be nulled. Using finite-element analysis, a depth of cut at the vertices is found for which the length change, on application of a compressive force, goes to zero and this is shown in Figure~\ref{fig2}.

Frictional forces at the supports mean that the cavity is, in fact, constrained in 12 degrees of freedom, 6 of them corresponding to rigid-body motion arising from inertial forces and 6 corresponding to deformation arising from differential forces. For deformations where the supports are forced towards each other in any one of the directions defined by the edges of the cube, the length change on axis is nulled by virtue of the truncated geometry. For deformations where two supports are forced towards each other and two are forced away from each other, the axial displacements at the mirror centres are either zero or cancel out and there is no length change on axis. Thus, the cavity is insensitive to deformations arising from differential forces for all degrees of freedom.

The cubic spacer is made from ultra-low expansivity glass and has an edge dimension of 50\,mm. The vertices are truncated to a depth of 6.7\,mm towards the centre of the spacer. Three cylindrical bores of 5.1\,mm diameter pass through the face-centres of the cube and allow for evacuation of the cavity. The silica mirrors are 12.7\,mm in diameter, 4.0\,mm thick and have a concave radius of curvature of 0.5\,m. They are optically contacted to the spacer and the cavity has a finesse in excess of 300,000. The cavity is housed within a vacuum chamber which also has a cubic geometry. The supports are made from nylon and are hemispherical with a diameter of 12.7\,mm. These are rigidly attached to posts mounted from vertex flanges of the chamber. The position of the cavity is gauged relative to the chamber so that it is placed symmetrically at its centre to within $100\,\mu\mathrm{m}$. A compressive force of up to 100\,N is then applied to the cavity by tightening the bolts for the flanges from which the supports are mounted. The chamber is supported between two breadboards on which the various optics and electronics for beam delivery and light detection are mounted. The setup is sufficiently compact and robust that it can be manipulated into any orientation. Light at 1064\,nm from a Nd:YAG laser, mounted on a separate platform, is transmitted to the cavity via an optical fibre and is frequency-locked to the cavity.

\begin{figure}[htb]
\centerline{\includegraphics[width=7.5cm]{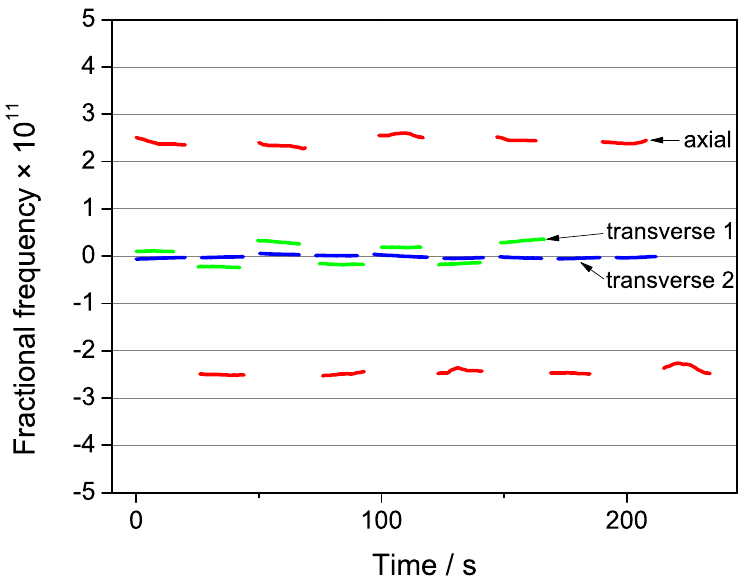}}
\caption{Fractional frequency as a function of time on repeated inversion. Linear drift is subtracted from the data and transients which occur during inversion are removed for clarity. \label{fig3}}
\end{figure}

To measure its sensitivity to inertial force, the cavity is inverted. This reverses the sign of gravity in the frame of the cavity and is equivalent to a change in acceleration of $2\mathit{g}$. The frequency of the laser locked to the test cavity is compared, by means of a heterodyne beat measurement, to that of a second laser locked to a reference cavity. The cavity is repeatedly inverted while the frequency difference is recorded on a counter. In between inversions, the cavity remains static for a period of around 25\,s. The laser remains locked to the cavity throughout. The measurement is repeated for the axial and two transverse directions which form an orthogonal set. Figure~\ref{fig3} shows the results.

Steps are observed corresponding to the reversals in the sign of gravity and the frequency alternates between two levels which correspond to the periods in between inversions when the cavity is static. The acceleration sensitivities are $2.45(3)/0.21(4)/0.01(1)\times 10^{-11}/\mathit{g}$ in the axial and two transverse directions respectively. The maximum acceleration sensitivity is $2.5\times 10^{-11}/\mathit{g}$.

The residual sensitivity is due to inaccuracy in realizing perfect symmetry. By a simple analysis, an offset in the position of the cavity relative to the chamber and supports of $\epsilon=100\mu\mathrm{m}$, results in a sensitivity of $(\rho\epsilon/E)\times 9.81= 3.2\times 10^{-11}/\mathit{g}$, where $\rho=2210\,\mathrm{kg}\,\mathrm{m}^{-3}$ and $E=67.6\,\mathrm{GPa}$ are the density and Young's modulus of the spacer material. The errors in the alignment of the mirrors with respect to the spacer and the centration of the curved mirror surfaces with respect to the mirror substrates are measured to be $\leq 60\,\mu\mathrm{m}$ and $\leq 15\,\mu\mathrm{m}$ respectively. Finite-element analysis shows that this leads to an axial(transverse) sensitivity of $\leq 0.1(0.3)\times 10^{-11}/\mathit{g}$. The sensitivity in the axial direction is likely to be due to error in the alignment of the cavity with respect to the chamber. This alignment is better in the transverse directions and the observed sensitivities may be due, in part, to misalignment of the mirrors.

In summary, a force-insensitive optical cavity has been designed for use in a non-laboratory environment. A cubic geometry is chosen which realizes the highest degree of symmetry between a four-point tetrahedral support and the optical axis. The cavity is constrained in all degrees of freedom and is insensitive to inertial forces acting in any direction. The vertices of the cube are truncated and this provides a means of nulling the sensitivity of the cavity to the compressive force by which it is held. In a robust experimental setup, the cavity operates in any orientation and a laser remains locked to it during inversion. The maximum acceleration sensitivity is $2.5\times 10^{-11}/\mathit{g}$ and this is the lowest reported passive sensitivity for an optical cavity. The residual sensitivity is accounted for by error in the alignment of the cavity with respect to its supports. With an improved gauge of position, it may be possible to reduce the passive sensitivity and a lower effective sensitivity could be achieved through use of active feed-forward correction for acceleration.

This work was supported by the NPL Strategic Research Program and by the European Space Agency under contract AO/1-5972/08/NL/CP.

\end{document}